\documentclass[showpacs,prc,preprint,nofootinbib,showkeys]{revtex4}
\pdfoutput=1

\usepackage{graphicx}
\usepackage{amsmath}
\usepackage{amssymb}
\usepackage{bm}
\usepackage{epstopdf}
\usepackage[colorlinks=true,linktocpage=true,linkcolor=blue,citecolor=blue]{hyperref}
\usepackage[usenames,dvipsnames]{color}

\newcommand{\beq}{\begin{equation}}
\newcommand{\eeq}{\end{equation}}
\newcommand{\bqa}{\begin{eqnarray}}
\newcommand{\eqa}{\end{eqnarray}}

\newcommand{\ph}{\hat{p}}
\newcommand{\pho}{\hat{p}_\Omega}
\newcommand{\phv}{\hat{p}_\varsigma}
\newcommand{\xh}{\hat{x}}

\newcommand{\atanh}{\tanh^{-1}}

\begin{document}


\title{Investigating the domain of validity of the Gubser solution to the Boltzmann equation}
  
\author{Ulrich Heinz}
\affiliation{Department of Physics, The Ohio State University,
  Columbus, OH 43210 United States}

\author{Mauricio Martinez}
\affiliation{Department of Physics, The Ohio State University,
  Columbus, OH 43210 United States}

\begin{abstract}
We study the evolution of the one particle distribution function that solves exactly the relativistic Boltzmann equation within the relaxation time approximation for a conformal system undergoing simultaneously azimuthally symmetric transverse and boost-invariant longitudinal expansion. We show, for arbitrary values of the shear viscosity to entropy density ratio, that the distribution function can become negative in certain kinematic regions of the available phase space depending on the boundary conditions. For thermal equilibrium initial conditions, we determine numerically the physical boundary in phase space where the distribution function is always positive definite. The requirement of positivity of this particular exact solution restricts its domain of validity, and it imposes physical constraints on its applicability.  
\end{abstract}


\date{\today }

\maketitle 


\section{Introduction}
\label{sec:intro}

The dynamics and transport properties of rarefied gases and fluids are usually described in terms of the Boltzmann equation. The Boltzmann equation is a partial integro-differential equation for the distribution function $f(x,p)$. In general this equation is solved numerically, and very few exact solutions are known in the literature. Nevertheless, for highly symmetric systems it is possible to solve this equation analytically under certain approximations for the collisional kernel~\cite{Cercignani2}. Using a relaxation time approximation (RTA) for the collisional kernel, the relativistic Boltzmann equation has been solved exactly for systems undergoing Bjorken flow~\cite{Baym:1984np} and Gubser flow~\cite{Denicol:2014tha,Denicol:2014xca}. The exact solution for the Bjorken flow has been useful to understand aspects of the isotropization and thermalization problem of a plasma formed by quarks and gluons (QGP)~\cite{Baym:1984np,Florkowski:2013lza,Florkowski:2013lya,Florkowski:2014sfa,
Denicol:2014tha,Denicol:2014xca,Nopoush:2014qba,Hatta:2015kia,Noronha:2015jia}. 
In contrast to Bjorken flow~\cite{Bjorken:1982qr}, where the system expands in boost invariant fashion only along one direction (the ``longitudinal" direction), Gubser flow~\cite{Gubser:2010ze,Gubser:2010ui} describes systems that undergo additionally simultaneous azymuthally symmetric expansion in the transverse directions. The solution of the Boltzmann equation for the Gubser flow~\cite{Denicol:2014tha,Denicol:2014xca} was found by exploiting the $SO(3)\otimes SO(1,1)\otimes Z_2$ symmetry of the flow velocity profile~\cite{Gubser:2010ze,Gubser:2010ui} which becomes manifest when mapping Minkowski space, $R^3\otimes R$, conformally onto de Sitter space times a line, $dS_3\otimes R$~\cite{Gubser:2010ze,Gubser:2010ui}. In Refs.~\cite{Denicol:2014tha,Denicol:2014xca} it was noticed that the resulting solutions for moments of the distribution function, such as the energy density or temperature, can become complex and therefore physically meaningless when propagating backwards in the de Sitter time (see discussion in Appendix B of Ref.~\cite{Denicol:2014xca}). In this work we revisit this issue and find that the unphysical behavior of the moments of the distribution function found in Ref.~\cite{Denicol:2014xca} is rooted in a violation of the positivity of the distribution function in some regions of phase space when propagating the solution of the Boltzmann equation for equilibrium initial conditions backward in the de Sitter time. In Minkowski coordinates, this translates to the distribution function becoming negative in certain momentum ranges at the outer edge of the spatial density profile at fixed time. 

This work is organized as follows: in Sect.~\ref{sec:exact-sol} we review the procedure used in Refs.~\cite{Denicol:2014tha,Denicol:2014xca} to find the exact solution of the Boltzmann equation~\cite{Denicol:2014tha,Denicol:2014xca}. In Sect.~\ref{sec:results} we show numerical results for the phase space evolution of the distribution function. Our conclusions are presented in Sect.~\ref{sec:concl}.

\section{The analytical solution of the RTA Boltzmann equation for the Gubser expansion}
\label{sec:exact-sol}

In this section we briefly review the derivation of the exact solution to the RTA Boltzmann equation that is invariant under the group of symmetries of the Gubser flow, i.e., under the $SO(3)_q\otimes SO(1,1)\otimes Z_2$ group (``Gubser group")~\cite{Gubser:2010ze,Gubser:2010ui}. For additional technical details of the method discussed here we refer the reader to Ref.~\cite{Denicol:2014xca}.

We use the following notation. The metric signature is taken to be the ``mostly plus'' convention. In Minkowski space the line element is written in Milne coordinates $x^{\mu }=(\tau, r, \phi, \varsigma)$ as
\begin{equation}
ds^{2}=g_{\mu \nu }dx^{\mu }dx^{\nu }=-d\tau^{2}+dr^{2}+r^2d\phi^{2}+\tau^2 d\varsigma^{2}\,.
\end{equation}
where the longitudinal proper time $\tau$, the spacetime rapidity $\varsigma$, the transverse radius $r$ and the azymuthal angle $\phi$ are defined in terms of the usual cartesian coordinates ($t,x,y,z$) as  
\begin{equation}
\begin{split}
&\tau =\sqrt{t^{2}{-}z^{2}}\,,\hspace{2cm}\varsigma =\atanh{\left(\frac{z}{t}\right)}\\
&r=\sqrt{x^{2}{+}y^{2}}\,,\hspace{2cm}\phi=\arctan{\left(\frac{y}{x}\right)}\,.
\end{split}
\end{equation}
The flow velocity $u^{\mu }$ is normalized as $u_{\mu }u^{\mu }=-1$. 

\subsection{The Gubser flow}
\label{subsec:Gflow}

The dynamics of an expanding conformal fluid in Minkowski space can be understood in terms of a  static conformal fluid defined in a particular curved space. In Minkowski space, the Gubser flow describes a system which expands azymuthally symmetrically in the transverse plane and at the same time in a boost invariant manner along the longitudinal direction. After applying a conformal map between Minkowski space and the de Sitter space times a line, $dS_3\otimes R$, the Gubser fluid velocity becomes static in this curved space. Thus, it is easiest to first study the dynamics in de Sitter space, and then map the solution back to Minkowski space~\cite{Gubser:2010ze,Gubser:2010ui}. 

In the de Sitter space, the Gubser symmetry can be made manifest by a suitable choice of coordinates. One relates the Milne coordinates $x^\mu=(\tau,r,\phi,\varsigma)$ defined in Minkowski space to a coordinate system $\xh^\mu=(\rho,\theta,\phi,\varsigma)$ in $dS_3\otimes R$ as follows~\cite{Gubser:2010ui}: One first performs a Weyl rescaling of the metric:
\begin{equation}
d\hat{s}^2=\frac{ds^2}{\tau^2}
=\frac{-d\tau^2+dr^2+r^2 d\phi^2}{\tau^2}+d\varsigma^2\,.  
\label{metricdS3R}
\end{equation}
Next, we make the following change of variables:
\begin{subequations}
\label{eq:rhotheta}
\begin{align}
\rho(\tilde\tau,\tilde r)& =-\mathrm{arcsinh}\left( \frac{1-\tilde\tau^2+\tilde r^2}
{2\tilde\tau }\right)\,,\hspace{1cm}   \rho\in (-\infty,\infty)\,,
\label{definerho} \\
\theta (\tilde\tau,\tilde r)& =\mathrm{arctan}\left(\frac{2\tilde r}{1+\tilde\tau^2-\tilde r^2}\right) \,,\hspace{2cm}   \theta\in (0,2\pi)\,, 
\label{definetheta}
\end{align}
\end{subequations}
where $\tilde\tau = q\tau$ and $\tilde r = q r$ with $q$ being an energy scale which sets the typical transverse size of the system~\cite{Gubser:2010ze,Gubser:2010ui}. Substituting Eq.~\eqref{eq:rhotheta} into Eq.~\eqref{metricdS3R} one finds  
\begin{equation}
d\hat{s}^{2}=-d\rho ^{2}+\cosh ^{2}\!\rho \left( d\theta ^{2}+\sin
^{2}\theta\, d\phi ^{2}\right) +d\varsigma ^{2}\,.  \label{eq:linedS3R}
\end{equation}
In these coordinates the metric in the curved $dS_3\otimes R$ space is then given by $\hat{g}_{\mu\nu}=\mathrm{diag}(-1,\cosh^2\rho,\cosh^2\rho\,\sin^2\theta,1)$. Eq.~\eqref{eq:linedS3R} shows that the variable $\rho$ is a time-like variable.
 
The Gubser flow is defined in $dS_3\otimes R$ as the unit vector $\hat{u}^\mu=(1,0,0,0)$. It is the only time-like unit vector which is invariant under the generators of the Gubser symmetry group~\cite{Gubser:2010ze,Gubser:2010ui}. The fluid velocity in Minkowski space is obtained by going back from the coordinates $\xh^\mu$ in $dS_3\otimes R$ to the coordinates $x^\mu$ in $R^3\otimes R$. This involves a Weyl rescaling of the fluid velocity components~\cite{Gubser:2010ze,Gubser:2010ui}. One finds that in Milne coordinates the fluid velocity in Minkowski space is $u^\mu=(\cosh\kappa(\tilde\tau,\tilde r),\sinh\kappa(\tilde \tau,\tilde r),0,0)$, with the transverse flow rapidity ~\cite{Gubser:2010ze,Gubser:2010ui}
\beq
\kappa(\tilde\tau,\tilde r)=\atanh\left(\frac{2\tilde\tau\tilde r}{1+\tilde\tau^2+\tilde r^2}\right)\,.
\eeq
This gives rise to the radial velocity profile 
\beq
\label{eq:velrad}
v(\tilde\tau,\tilde r) = \tanh\kappa(\tilde\tau,\tilde r)=\frac{2\tilde\tau\tilde r}{1+\tilde\tau^2+\tilde r^2}\,.
\eeq
 
\subsection{The exact solution of the RTA Boltzmann equation}
\label{subsec:exact}

The invariance of a system under a particular group of transformations imposes constraints on the number of independent variables of the distribution function $f(x^\mu,p_i)$. For the Gubser group one finds that $f(\xh^\mu;\ph_i)={f}(\rho;\hat{p}_\Omega^2,\hat{p}_\varsigma)$~\cite{Denicol:2014tha,Denicol:2014xca} where  $\ph^2_\Omega=\ph_\theta^2+\ph_\phi^2/\sin^2\theta$ and $\ph_\theta$, $\ph_\phi$ and $\ph_\varsigma$ are the momenta conjugate to the coordinates $\theta$, $\phi$ and $\varsigma$ in Eq.~\eqref{eq:linedS3R}. Thus, the relativistic RTA Boltzmann equation in de Sitter space reduces to a one-dimensional ordinary differential equation in de Sitter space~\cite{Denicol:2014tha,Denicol:2014xca},
\beq
\frac{\partial}{\partial \rho}f(\rho;\hat p_\Omega^2,\hat p_\varsigma)= -\frac{\hat{T}(\rho)}{c}\left[f(\rho;\hat p_\Omega^2,\hat p_\varsigma)-f_{\rm eq}\left(\frac{\hat p^\rho}{\hat T(\rho)}\right)\right]\,,
\label{newRTAboltzmanneq}
\eeq
where $\ph^\rho=\ph^\rho(\rho,\ph_\Omega,\ph_\varsigma)=\sqrt{\ph^2_\Omega/\cosh^2\rho+\ph^2_\varsigma}$~, $f_{eq}(x)=e^{-x}$ is the Boltzmann equilibrium distribution function, and $\hat{T}=c/\hat{\tau}_{rel}(\rho)$ and $\hat{\tau}_{rel}(\rho)=\tau_{rel}(\tau)/\tau$ are the Weyl-rescaled (unitless) temperatures and local relaxation time. The constant $c$ is related to the specific shear viscosity of the system by $c=5\eta/{\mathcal S}$~\cite{Denicol:2010xn,Denicol:2011fa,Florkowski:2013lza,Florkowski:2013lya} where $\eta$ is the shear viscosity and ${\mathcal{S}}$ is the entropy density. The formal solution of this equation is~\cite{Cercignani,Cercignani2,Baym:1984np}
\beq
f(\rho;\hat p_\Omega^2,\hat p_\varsigma) =D(\rho,\rho_0) f_0(\rho_0;\hat p_\Omega^2,\hat p_\varsigma)\,+\frac{1}{c}\int_{\rho_0}^\rho d\rho'\,D(\rho,\rho')\,\hat T(\rho')\, f_{\rm eq}(\ph^\rho/\hat{T}(\rho)) \, ,
\label{boltzmannsolution}
\eeq
where $D(\rho,\rho_0)=\exp\!\left[-\int_{\rho_0}^\rho d\rho' \,\hat T(\rho')/c\right]$ is the damping function, and  $f_0(\rho_0;\hat p_\Omega^2,\hat p_\varsigma)$ is the initial distribution function at $\rho_0$ which we choose to be the Boltzmann equilibrium distribution, $f_0(\rho_0,\ph^2_\Omega,\ph_\varsigma)=f_{eq}\left(\ph^\rho(\rho_0)/\hat{T}(\rho_0)\right)$. 

The temperature $\hat{T}$ is related to the solution for $f$ by the dynamical Landau matching condition, i.e., by the requirement that $\hat{\varepsilon}(\rho)=\hat{\varepsilon}_{eq}(\rho)\sim \hat{T}^4(\rho)$ where $\hat{\varepsilon}(\rho)$ is the energy density computed from the non-equilibrium distribution function $f(\rho,\ph_\Omega^2),\ph_\varsigma$. This matching condition allows to rewrite Eq.~\eqref{boltzmannsolution} as~\cite{Denicol:2014tha,Denicol:2014xca}
\begin{equation}
\hat{T}^{4}(\rho )=D(\rho ,\rho _{0})\mathcal{H}
\left( \frac{\cosh \rho _{0}}{\cosh\rho}\right) \hat{T}^{4}(\rho _0)
+\frac{1}{c}\int_{\rho_0}^{\rho}d\rho'\,D(\rho,\rho')\,
\mathcal{H}\left(\frac{\cosh\rho'}{\cosh\rho}\right) \,\hat{T}^{5}(\rho')\,,
\label{eq:efftemp}
\end{equation}
where
\beq
\mathcal{H}(x)=\frac{1}{2}\left(x^2+x^4\frac{\tanh^{-1}\left(\sqrt{1-x^2}\right)}{\sqrt{1-x^2}}\,\right)\,.
\eeq
The numerical solution of~\eqref{eq:efftemp} is uniquely determined by the initial de Sitter time $\rho_0$, the initial value $\hat{T}_0=\hat{T}(\rho_0)$, and (through $c$) the chosen value of $\eta/{\mathcal S}$. In order to solve for $f$ in Eq.~\eqref{boltzmannsolution}, we first find the temperature $\hat{T}(\rho)$ by iteratively solving the integral equation~\eqref{eq:efftemp} as in~\cite{Florkowski:2013lza,Florkowski:2013lya}, and then plug this $\hat{T}(\rho)$ into Eq.~\eqref{boltzmannsolution} for every de Sitter time-step $\rho$, and perform the $\rho'$ integral on the r.h.s. Since Eq.~\eqref{boltzmannsolution} is diagonal in the momentum variables $\hat{p}_\Omega$, $\hat{p}_\varsigma$, the evolution of $f$ with $\rho$ can be studied separately for each point in momentum-space. In Ref.~\cite{Denicol:2014tha,Denicol:2014xca} we showed that all macroscopic moments of the distribution function $f$, (i.e., all the components of the energy-momentum tensor $\hat{T}^{\mu\nu}$) can be directly obtained as $\rho'$-integrals over the solution $\hat{T}(\rho)$ from Eq.~\eqref{eq:efftemp}, with different weight functions, without even studying the distribution function $f$ in Eq.~\eqref{boltzmannsolution} itself.

In Ref.~\cite{Denicol:2014xca} (App.B) we observed for a few specific choices of $\rho_0$ that the solution of Eq.~\eqref{eq:efftemp} leads to complex-valued temperatures at large enough negative values of $\rho-\rho_0$. This observation was generic and held for any value of $\eta/{\mathcal S}$. The problem appeared to arise only for $\rho<\rho_0$, i.e., only in the de Sitter past; for $\rho>\rho_0$ the solution for $\hat{T}(\rho)$ was always real. Obviously, a complex-valued temperature is physically meaningless.  In practice this issue can be resolved by imposing boundary conditions in the infinitely distant past of the de Sitter time, i.e. for $\rho_0\to-\infty$~\cite{Denicol:2014xca}. It was argued in Ref.~\cite{Denicol:2014xca} that the complex values of the temperature might be related with the violation of the positivity condition of the distribution function in some regions of phase space. Clearly, if this condition is not satisfied, the probabilistic meaning of the distribution function is lost. 

In this work we investigate the evolution of the distribution function~\eqref{boltzmannsolution} in phase space for arbitrary values of $\rho_0$, $\hat{T}_0$ and $\eta/{\mathcal S}$. Results of our studies are presented in the following section. Before going into the discussion of our numerical results, though, we make the following analytical observation: Expanding the solution~\eqref{boltzmannsolution} around $\rho_0$ for small $\rho-\rho_0$ it is straightforward to see, for any choice of $(\ph_\Omega^2,\ph_\varsigma)$, that $f(\rho,\ph_\Omega^2,\ph_\varsigma)$ increases for $\rho>\rho_0$ and decreases for $\rho<\rho_0$. The rate of increase/decrease depends on $(\ph_\Omega^2,\ph_\varsigma)$. From this we see that, as the system evolves forward in de Sitter time $\rho$, $f$ will remain positive if it is so at $\rho=\rho_0$, but that $f$ will decrease, and can for any choice of $(\ph_\Omega^2,\ph_\varsigma)$ turn negative, as we follow the system backward in $\rho$ to sufficiently large negative values of $\rho-\rho_0$.

\section{Results}
\label{sec:results}
\begin{figure}[h]
\begin{centering}
\includegraphics[scale=0.2]{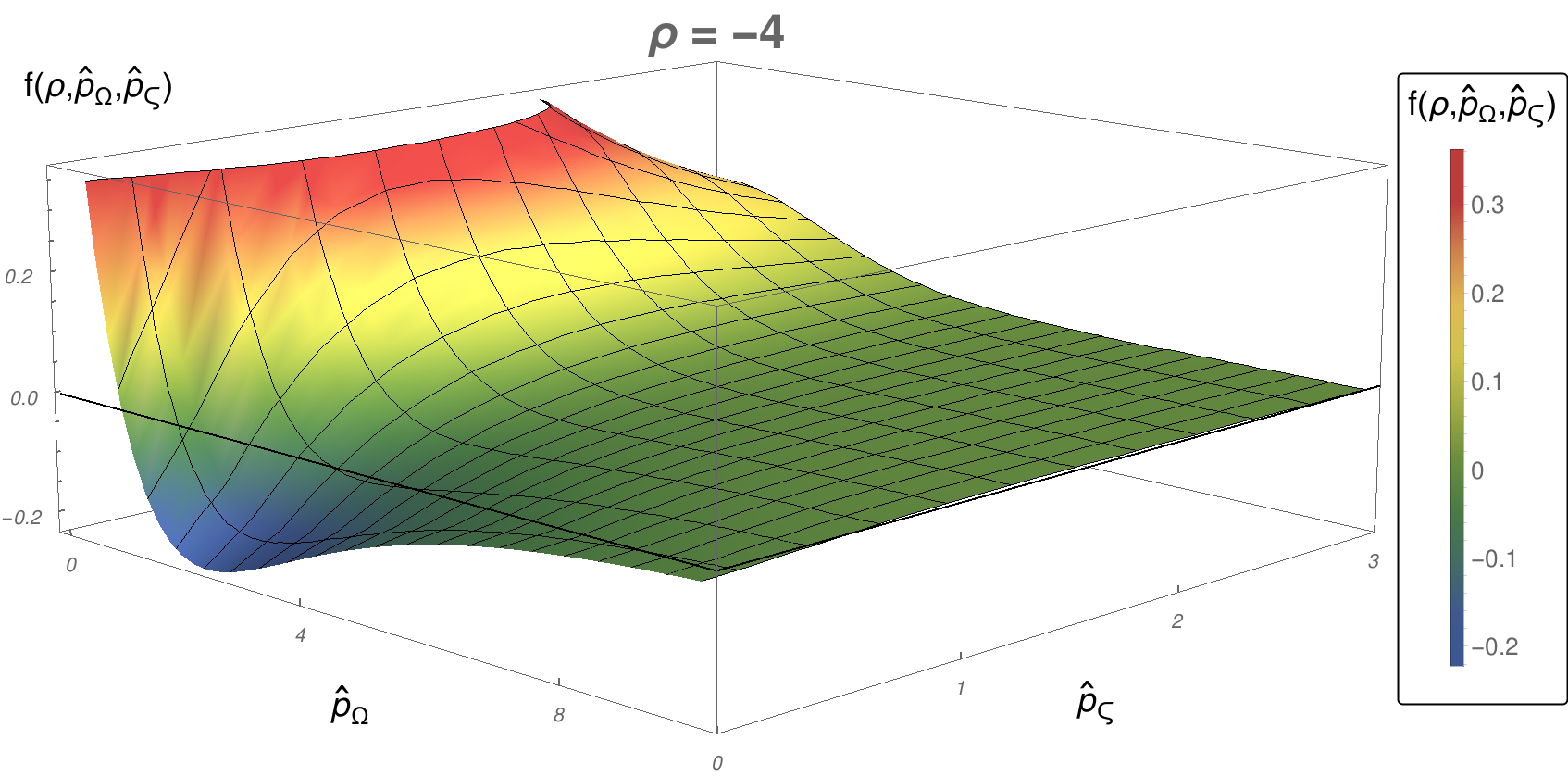}\vspace{0.5cm}
\includegraphics[scale=0.2]{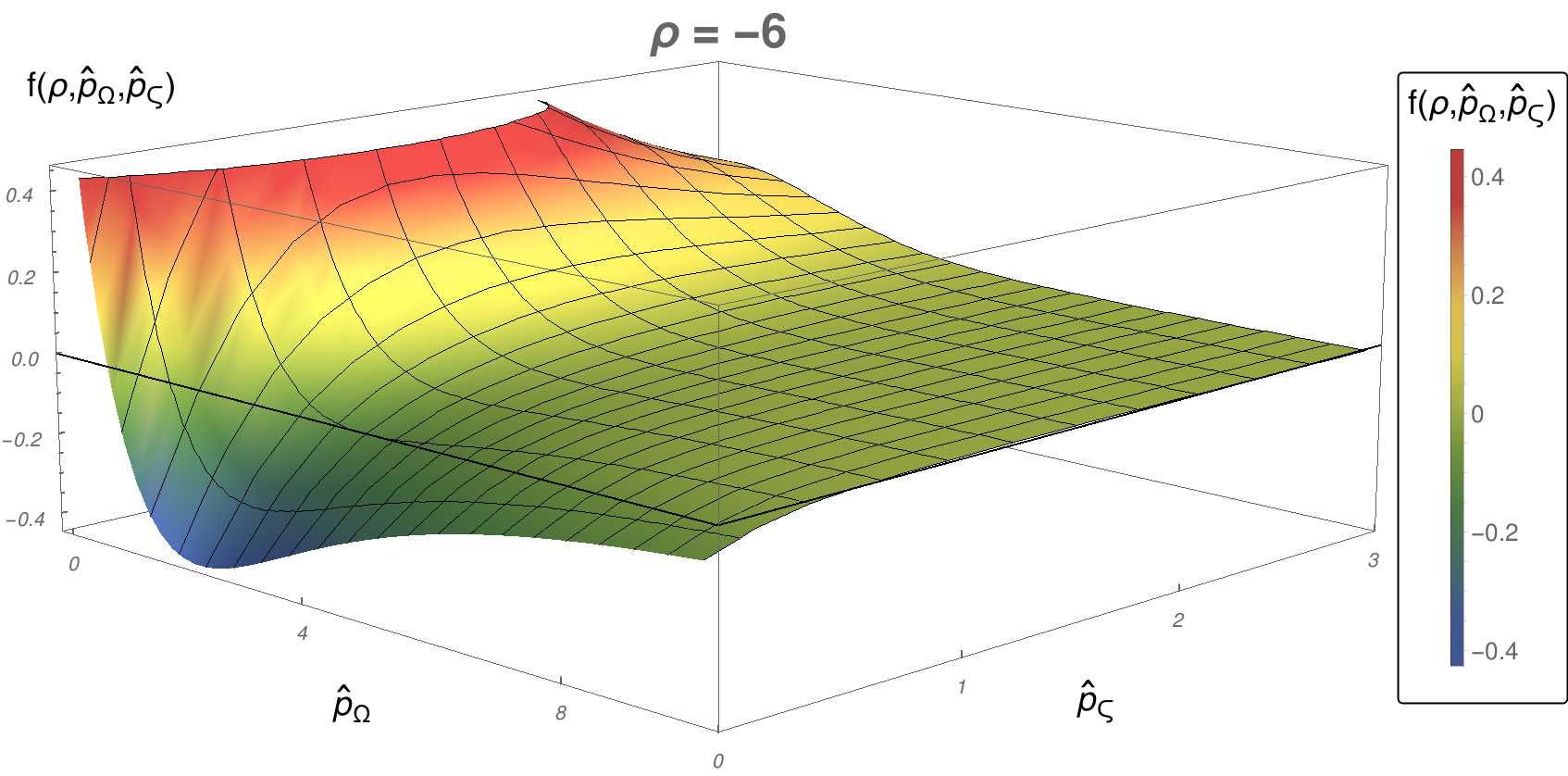}
\end{centering}
\caption{(Color online) Snapshots of the two-dimensional slice of the $\pho$ and $\phv$ evolution of the phase space distribution $f(\rho,\pho,\phv)$ as a function of $\ph_\Omega$ and $\ph_\varsigma$ for fixed values of $\rho=-4$ (top panel) and $\rho=-6$. In both figures we consider  $\rho_0=0$, $4\pi\eta/{\mathcal S} = 3$ and $\hat{T}_0=1$.} 
\label{F1}
\end{figure}
\begin{figure}[h]
\begin{centering}
\includegraphics[scale=0.2]{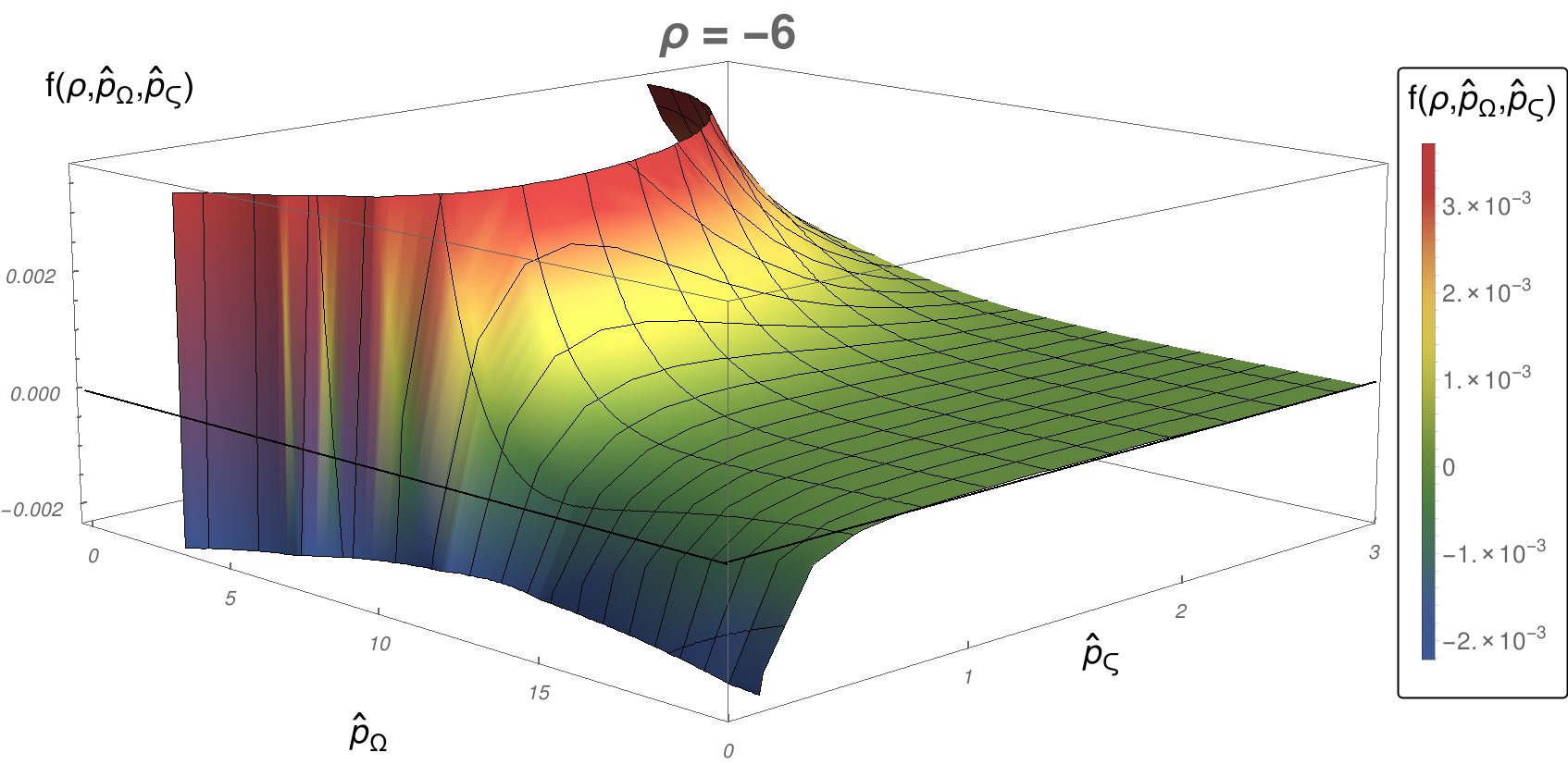}\vspace{0.5cm}
\includegraphics[scale=0.2]{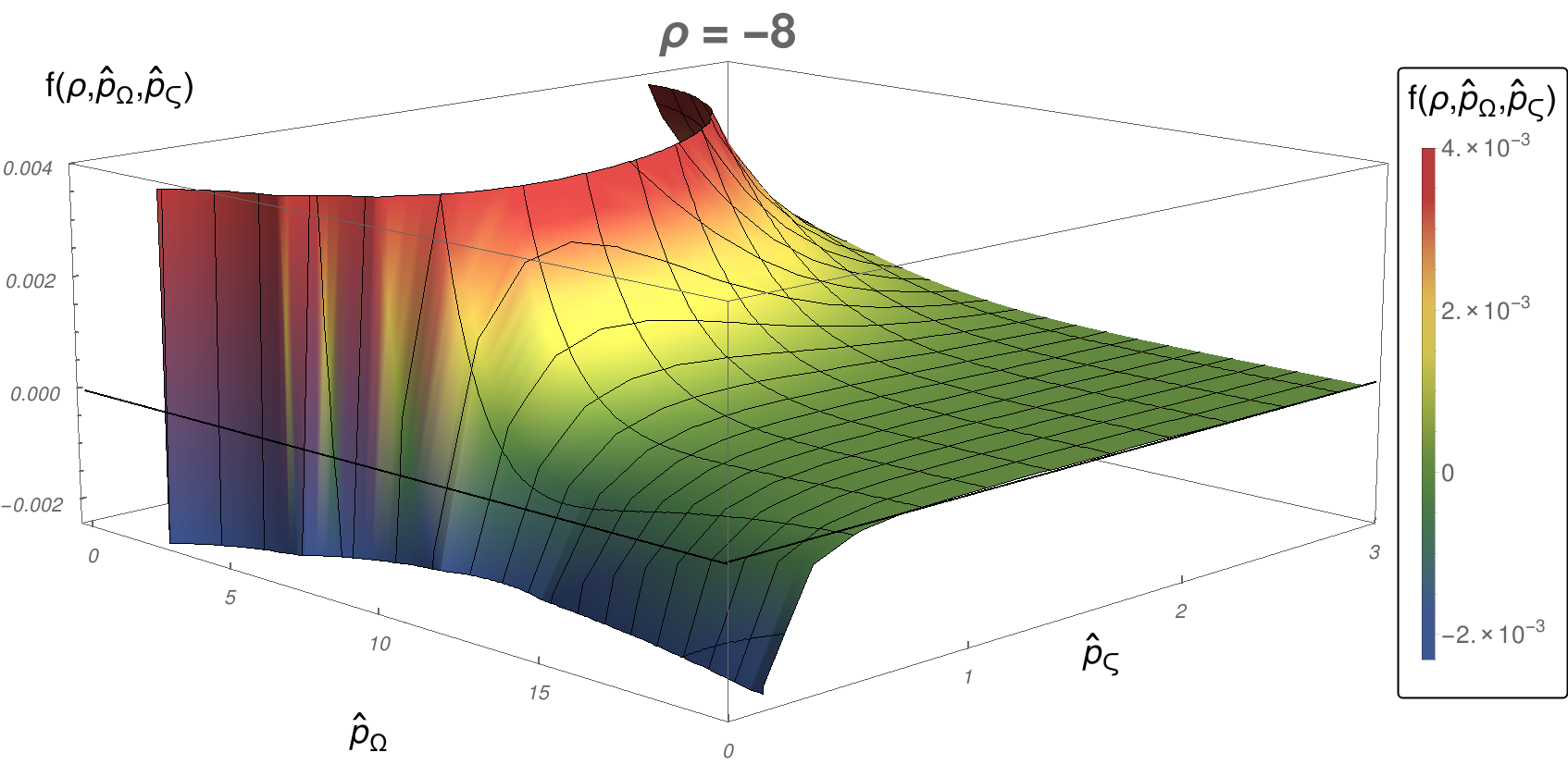}
\end{centering}
\caption{(Color online) Snapshots of the two-dimensional slice of the $\pho$ and $\phv$ evolution of the phase space distribution $f(\rho,\pho,\phv)$ as a function of $\ph_\Omega$ and $\ph_\varsigma$ for fixed values of $\rho=-6$ (top panel) and $\rho=-8$. In both figures we consider  $\rho_0=-2$, $4\pi\eta/{\mathcal S} = 3$ and $\hat{T}_0=1$.} 
\label{F2}
\end{figure}

\begin{figure}[t]
\begin{centering}
\includegraphics[scale=0.17]{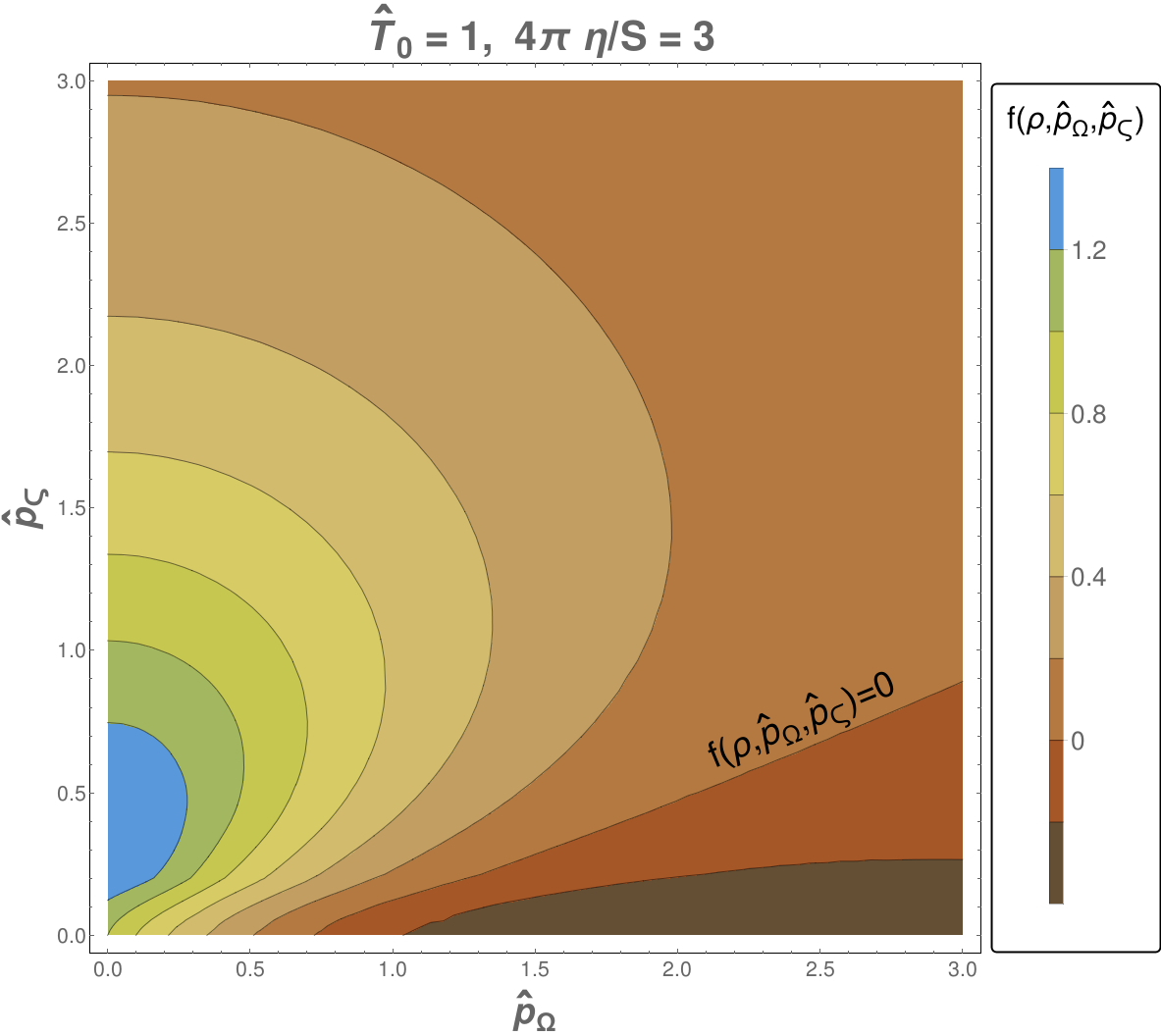}
\includegraphics[scale=0.17]{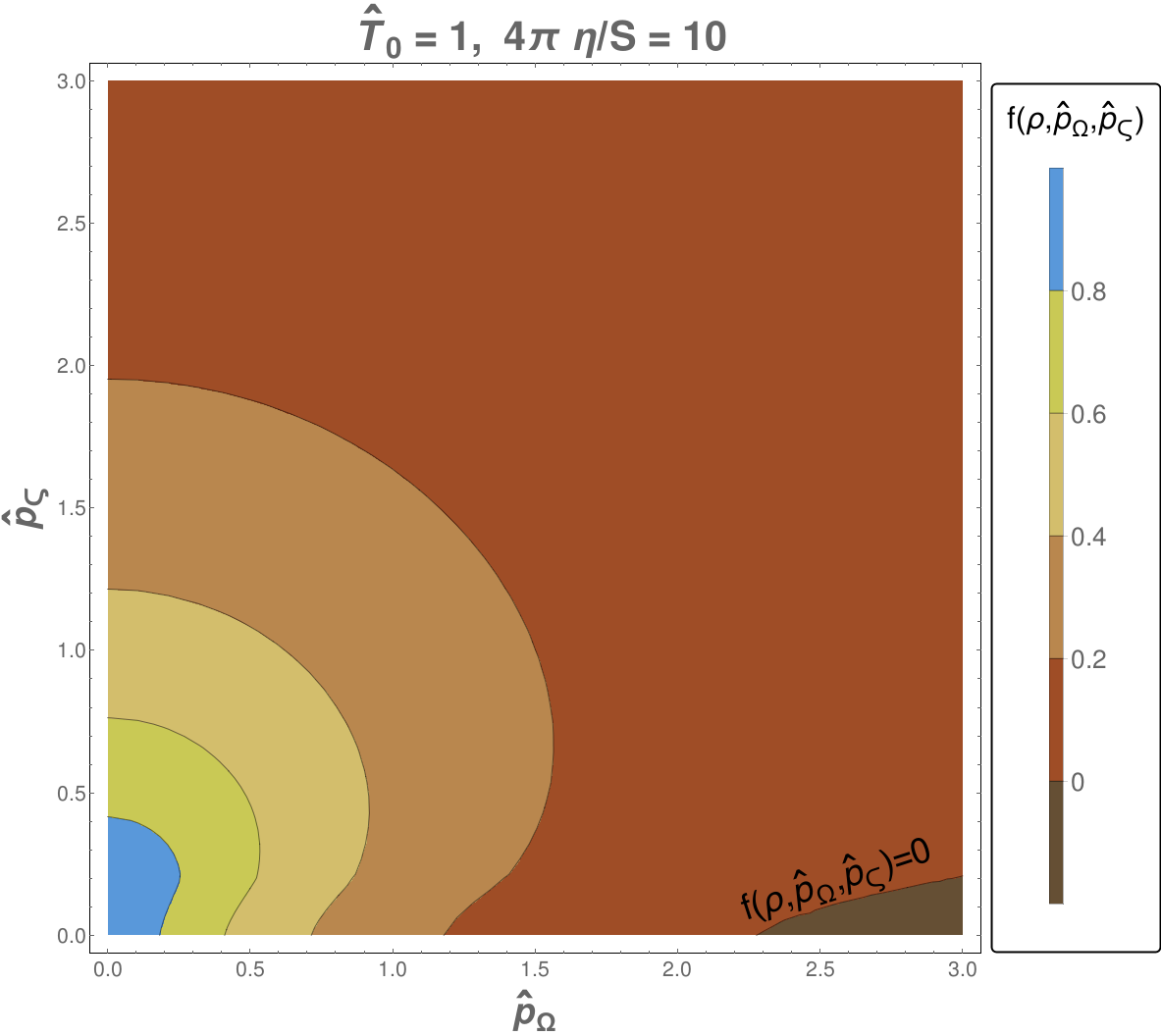}
\end{centering}
\caption{(Color online) Contour plots of $f(\rho,\ph^2_\Omega,\ph_\varsigma)$ for fixed values $\rho_0=0$ and $\rho=-6$ but different values of $(4\pi)\eta/{\mathcal S}=3$ (left panel) and $(4\pi)\eta/{\mathcal S}=10$ (right panel). In both panels $\hat{T}_0=1$.} 
\label{F3}
\end{figure}
\begin{figure}[t]
\begin{centering}
\includegraphics[scale=0.17]{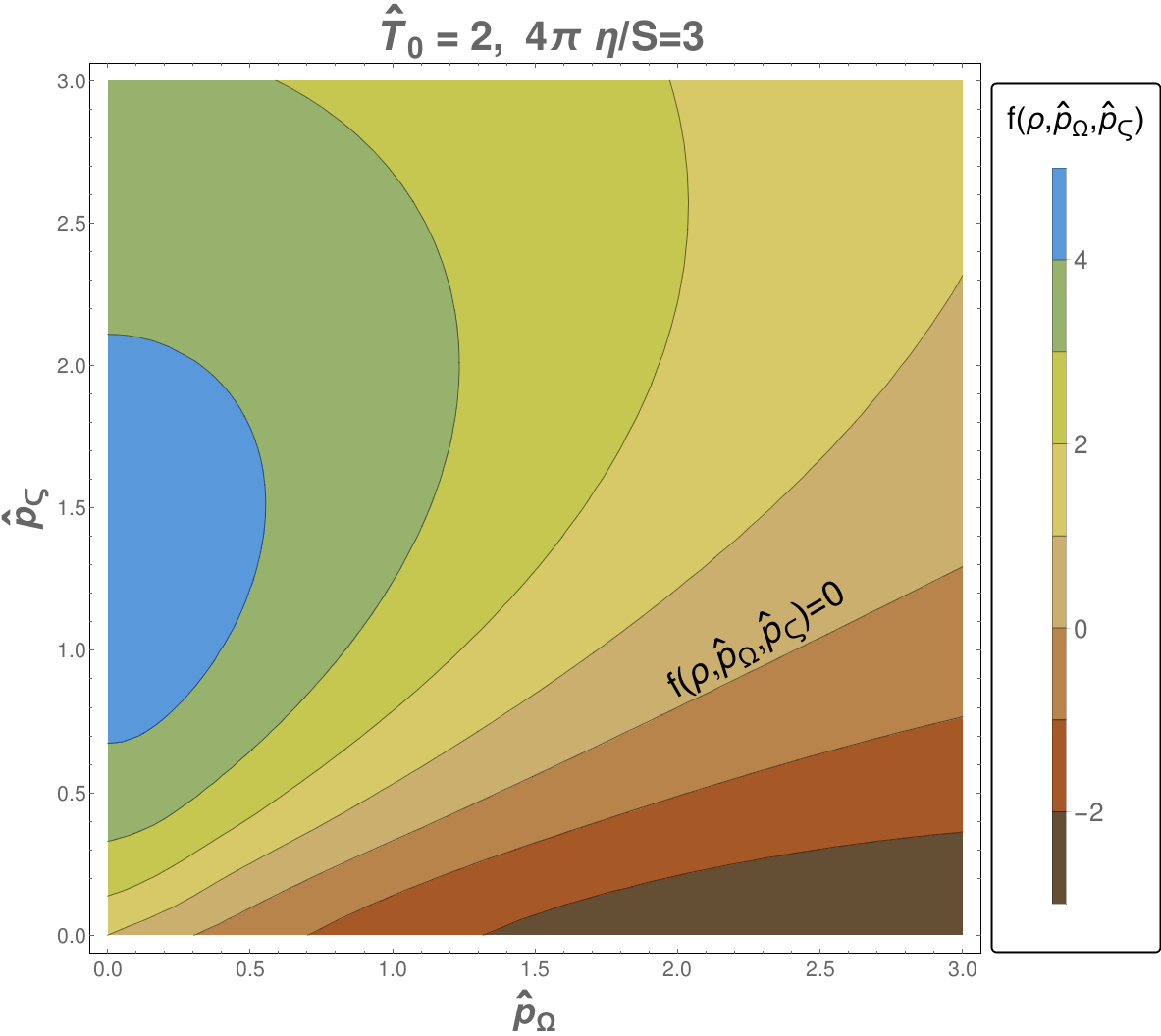}
\includegraphics[scale=0.17]{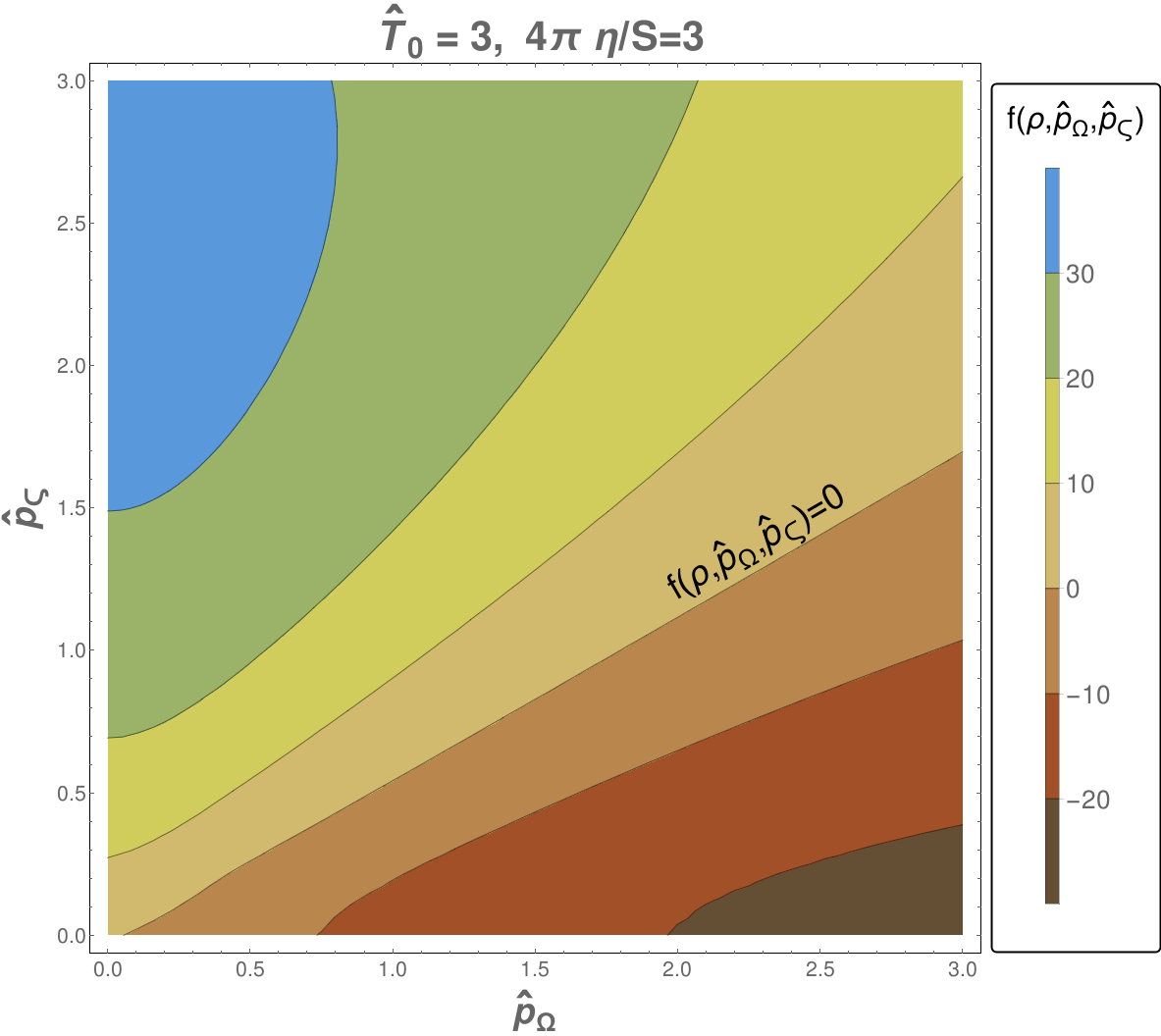}
\end{centering}
\caption{(Color online) Contour plots of $f(\rho,\ph^2_\Omega,\ph_\varsigma)$ for a fixed values of $\rho_0=0$ and $\rho=-6$ but different values of $\hat{T}_0=2$ (left panel) and $\hat{T}_0=3$ (right panel). In both panels $(4\pi)\eta/{\mathcal S}=3$.} 
\label{F4}
\end{figure}

In Fig.~\ref{F1} and~\ref{F2} we show contour plots of the distribution function $f(\rho,\ph_\Omega^2,\ph_\varsigma)$ as a function of $\ph_\Omega$ and $\ph_\varsigma$ for fixed values of the de Sitter time $\rho$, for the specific parameter choices $4\pi\eta/{\mathcal S}=3$ and $\hat{T}_0=1$. In Fig.~\ref{F1} we impose equilibrium boundary conditions at $\hat{\rho}_0=0$ and then study $f$ at $\rho=-4$ (top panel) and $\rho=-6$ (bottom panel). In Fig.~\ref{F2} we translate the entire problem by two units in de Sitter time $\rho$, imposing the same initial conditions at $\rho_0=-2$ and then studying $f$ at $\rho=-6$ (top panel) and $\rho=-8$ (bottom panel). Obviously, the solution is not translationally invariant in $\rho$. While in both figures we observe that the distribution function becomes negative at large values of $\hat{p}_\Omega$ and small values of $\hat{p}_\varsigma$, and that this problem becomes more severe as $|\rho-\rho_0|$ increases, i.e. as $\rho$ becomes more negative, the problem is clearly, for the same value of $|\rho-\rho_0|$, more serious for smaller initial values $\rho_0$.  

The fact that the distribution function goes negative in some region of phase space is independent of the value of $\eta/S$. This is seen in Fig.~\ref{F3} where we show $f=$ const. contours in the $\ph_\Omega-\ph_\varsigma$ plane for two different values of $\eta/{\mathcal{S}}$, namely $4\pi\eta/{\mathcal S}=3$ (left panel) and  $4\pi\eta/{\mathcal S}=10$ (right panel). In both panels we imposed equilibrium initial conditions with $\hat{T}_0=1$ at $\rho_0=0$ and plotted $f$ at $\rho=-6$. For fixed $|\rho-\rho_0|$, as $\eta/{\mathcal{S}}$ increases the line $f(\rho,\ph_\Omega^2,\ph_\varsigma)=0$ separating physical ($f>0$) from unphysical behavior ($f<0$)  is seen to move closer to the $\ph_\varsigma = 0$ axis and towards larger $\ph_\Omega$ values. In other words, for larger specific shear viscosity, the $\rho$-evolution that eventually drives $f$ negative proceeds more slowly. 

While the positivity condition of the phase space distribution is violated for any value of the temperature $\hat{T}_0$, the speed with which this happens as $\rho$ evolves backwards from $\rho_0$ depends strongly on the choice of $\hat{T}_0$ even if $\eta/\mathcal{S}$ is held constant.  This is perhaps not unexpected since $\hat{T}$ also controls the scattering rate (due to the conformal relation $\hat{T}\hat{\tau}=$const.). In Fig.~\ref{F4} we present contours of the distribution function in the $\ph_\Omega-\ph_\varsigma$ plane for two additional values of the initial temperature, $\hat{T}_0=2$ (left panel) and $\hat{T}_0=3$ (right panel) (in addition to the case $\hat{T}_0=1$ shown in Fig.~\ref{F3}), all for the same value $4\pi \eta/\mathcal{S}=3$. As in Fig.~\ref{F3}, equilibrium initial conditions are implemented at $\rho_0=0$, and $f$ is plotted for $\rho=-6$. As $\hat{T}_0$ is increased, the line $f=0$ separating the physical from the unphysical region moves to smaller $\ph_\Omega$ and larger $\ph_\varsigma$ values, i.e. the unphysical region grows. Increasing the initial temperature and energy density obviously speeds up the evolution towards unphysical behavior as de Sitter times evolves backward. 

\begin{figure}[t]
\begin{centering}
\includegraphics[scale=0.15]{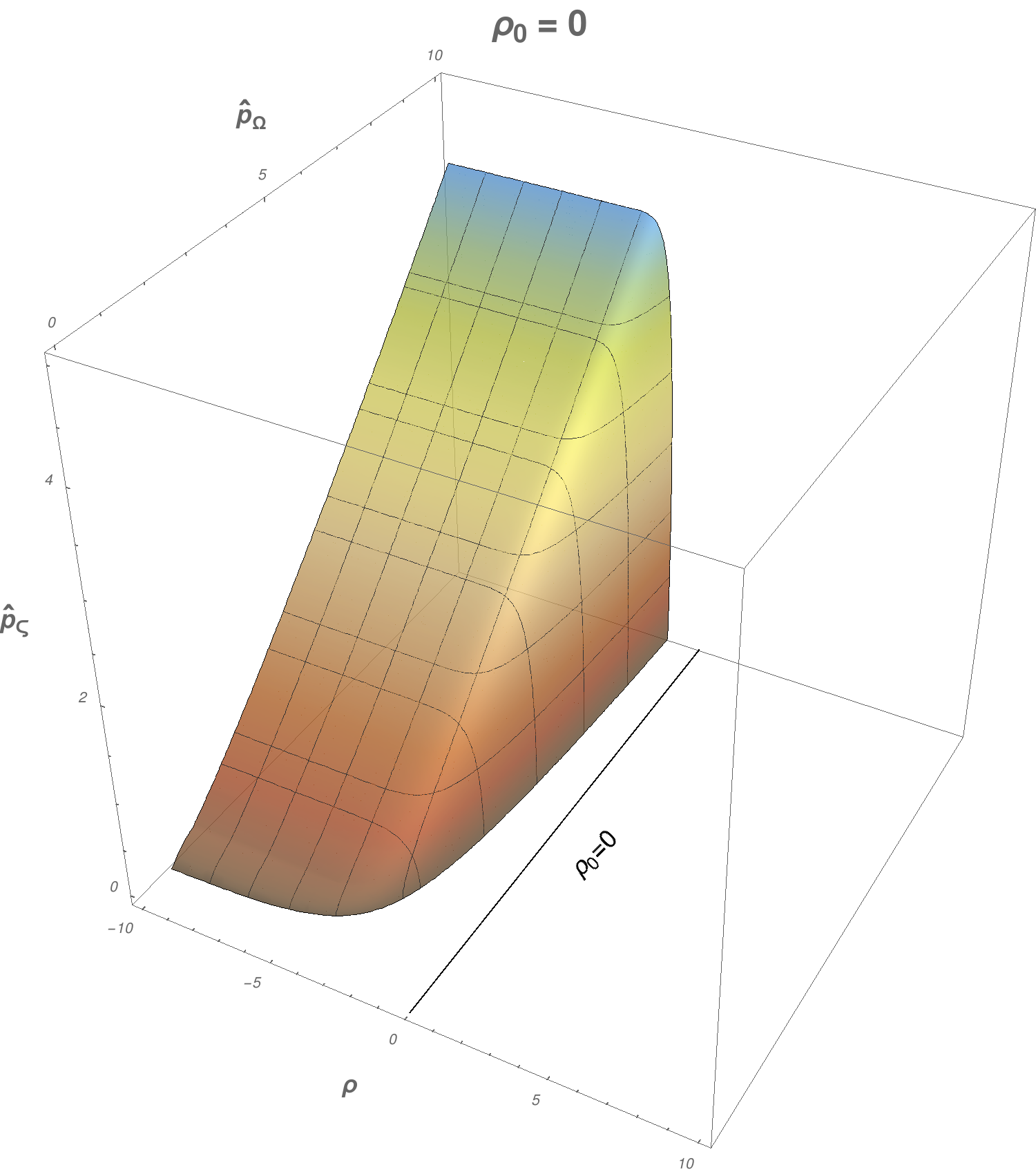}
\includegraphics[scale=0.15]{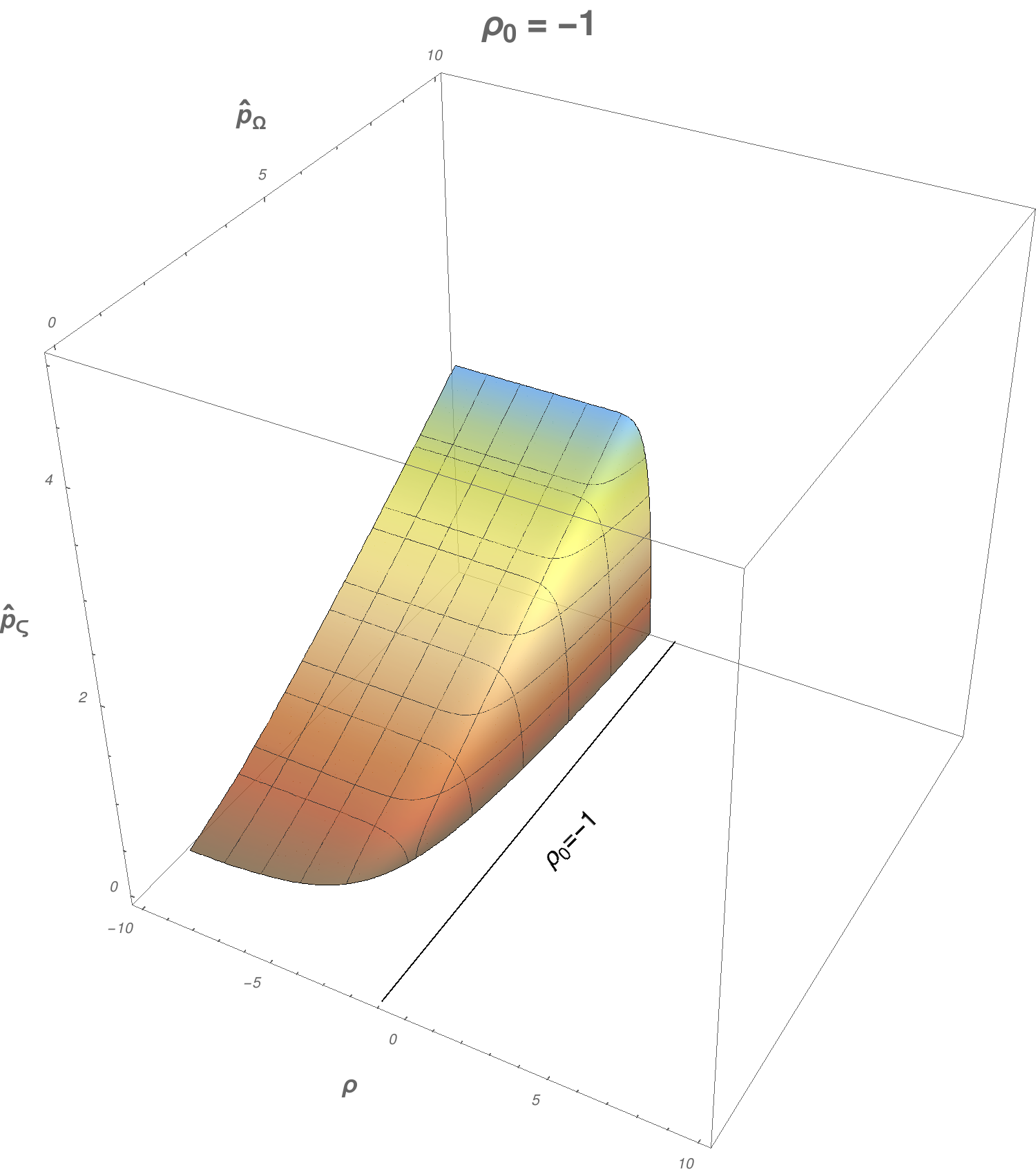}
\end{centering} 
\caption{(Color online) 3D surface defined by $f(\rho,\ph_\Omega,\ph_\varsigma)=0$ for two different initial conditions $\rho_0=0$ (left panel) and $\rho_0=-1$ (right panel). In both cases we chose $(4\pi)\eta/{\mathcal S}=3$ and $\hat{T}_0=1$. The gray lines drawn over the physical boundary condition $f=0$ correspond to constant values of $\rho-\ph_\Omega-\ph_\varsigma$ over the surface.} 
\label{F5}
\end{figure}

In Fig.~\ref{F5} we show this $\rho$ evolution by plotting the $f=0$ surface separating the physical from the unphysical regions in the 3-dimensional $\rho-\ph_\Omega-\ph_\varsigma$ space, for initial conditions imposed at $\rho_0=0$ (left panel) and $\rho_0=-1$, respectively. In this figure we do not show the region where $\ph_\varsigma < 0$ since the distribution function is invariant under reflexions of the  $\ph_\varsigma$ variable by construction~\cite{Denicol:2014tha,Denicol:2014xca}. Clearly the unphysical region only appears for $\rho < \rho_0$, first at large $\ph_\Omega$ and later (i.e. for more negative values of $\rho-\rho_0$), also for smaller $\ph_\Omega$ values, and it is largest when $\ph_\varsigma$ is small. As the initial time $\rho_0$ is decreased, the unphysical region appears more quickly as a function of $\rho$, but covers a smaller region in $\ph_\Omega$ and $\ph_\varsigma$.

The non-positive behavior of $f(\rho,\ph_\Omega,\ph_\varsigma)$ in the de Sitter space imposes limitations of the validity of this particular solution when the information is mapped back to Minkowski space. For massless particles the distribution function is a Lorentz and Weyl scalar~\cite{Denicol:2014xca}, so one only needs to transform the phase space coordinates $(\rho,\ph_\Omega,\ph_\varsigma)$ to the corresponding ones in Minkowski space $(\rho(\tau, r),p_\Omega^2 (\tau,r,p^\tau,p^r),p_\varsigma)$. When transforming the momentum coordinates $\ph^\mu$ in de Sitter to the corresponding ones in Minkowski space one must perform the associated Weyl rescaling ($\ph^\mu = \tau^2\,p^\mu$) together with the covariant transformation of the momentum components 
\beq
\begin{pmatrix}
\ph^\rho \\ \ph^\theta \\ \ph^\phi \\ \ph^\phi
\end{pmatrix}\,=\,
\begin{pmatrix}
\frac{\partial\rho}{\partial\tau} & \frac{\partial\rho}{\partial r} &0 & 0 \\
\frac{\partial\theta}{\partial\tau} & \frac{\partial\theta}{\partial r} &0 &0 \\
0 & 0 & 1 & 0 \\
0 & 0 & 0 &1
\end{pmatrix}
\,
\begin{pmatrix}
\tau^2\,p^\tau\\ \tau^2\,p^r \\ \tau^2\,p^\phi \\ \tau^2\,p^\varsigma
\end{pmatrix}\,. 
\eeq
By using the coordinate transformations~\eqref{eq:rhotheta} we obtain explicitly
\begin{subequations}
\label{eq:trans}
\begin{align}
\label{eq:transrho}
\ph^\rho&= \tau\,\gamma\,\left( p^{\,\tau} -
v(\tau, r)\, p^{\, r}
\right)\,,\\
\label{eq:transth}
\ph^\theta&=\frac{\tau}{\cosh\rho(\tau, r)}\,\gamma
\left( p^{\,r} -v(\tau,r)\, p^{\,\tau} 
\right) \,,\\
\ph^\phi&= \tau^2\,p^\phi \,,\\
\ph^\varsigma&=\tau^2\,p^\varsigma = p_\varsigma \,,
\end{align}
\end{subequations}
where $v(\tau, r)$ is the radial Gubser flow velocity in Minkowski space, given in Eq.~\eqref{eq:velrad}, and $\gamma=1/\sqrt{1-v^2}$. Eqs.~\eqref{eq:transrho} and ~\eqref{eq:transth} show that up to a Jacobian factor, related to the Weyl-rescaling, $(\ph^\rho,\ph^\theta)$ are related with the Minkowski-space components $(p^\tau,p^r)$ by a radial boost with velocity $v (\tau,r)$. Eqs.~\eqref{eq:trans} are inverted by
\begin{subequations}
\label{eq:transinv}
\begin{align}
p^\tau&= \frac{\gamma}{\tau}\,\left( \ph^{\rho} +v(\tau, r)\,\cosh\rho\,\ph^{\theta}
\right)\,,\\
p^r&=\frac{\gamma}{\tau}\,
\left(\cosh\rho\,\ph^{\theta} +v(\tau,r)\,\ph^{\rho} 
\right) \,,\\
p_\phi&=\,r^2\,p^\phi= \frac{r^2}{\tau^2}\,\ph^\phi \,,\\
p_\varsigma&=\,\ph^\varsigma \,,
\end{align}
\end{subequations}
Thus, we can understand the momentum $p^\mu$ in the lab frame in terms of the momentum in de Sitter space boosted by radial flow.

In order to develop some intuition what the break-down of the solution for the distribution function in de Sitter space implies in Minkowski space we make the following considerations:
\begin{itemize}
\item The scale of energy $q$ is determined by the transverse size $R_\perp$ of the nucleus; we assume $R_\perp =$ 5 fm and thus set $q= 1/R_\perp=$ 0.04 GeV. 
\item We study the system in Minkowski space at a fixed longitudinal proper time $\tau$ for which we take $\tilde\tau=q\tau=0.3$, i.e. $\tau$= 1.5 fm/$c$. 
\item The rotational invariance of the Gubser flow about the longitudinal axis allows us to choose $\ph_\phi=0$. Moreover, this lets us identify the radial component of the momentum in Milne coordinates with the transverse momentum of the particle.\footnote{In polar coordinates the momentum component $p^\phi$ is 
\begin{equation*}
p^\phi=\frac{p_T}{r}\sin (\phi-\phi_p)\,.
\end{equation*}
where $p_T$ is the transverse momentum of the particle and $\phi_p$ is the angle between the two-dimensional transverse momentum vector ${\bf p_T}$ and the horizontal axis in the transverse plane. Therefore fixing $p_\phi=0$ is equivalent to saying that the radial unit vector in the transverse plane is aligned with ${\bf p_T}$, i.e., $\phi = \phi_p$. It is straightforward to conclude that in this case $p^r=p_T$.}
\item $\ph_\varsigma=p_\varsigma$ is related to the longitudinal component of the momentum in the boosted frame, i.e. $p_\varsigma=p_z\,t-E\,z\equiv w$. It is an invariant under longitudinal boosts, i.e. under the subgroup $SO(1,1)$ of the Gubser group~\cite{Denicol:2014tha,Denicol:2014xca}. Thus we can evaluate $p_\varsigma$ at $z=0$ where $t$ and $\tau$ coincide and thus $p_\varsigma=\tau p_z$. 
\item For $z=0$ and $\ph_\phi=0$ we can write the $SO(3)_q$ invariant $\ph^2_\Omega$ as
\beq
\label{eq:simplepom}
\begin{split}
\ph^2_\Omega &=\ph^2_\theta +\frac{\ph_\phi^2}{\sin^2\theta}\,,\\
&= \left(\ph^\theta \cosh^2\rho(\tau,r)\right)^2\,\\
&=\cosh^2\rho(\tau, r)\,\tau^2\,\left[\gamma(\,p_T-v(\tau,r)\, p^\tau)\right]^2\,,
\end{split}
\eeq
where $p^\tau=\sqrt{p_T^2+p_z^2}$ and $p_T=p^r$.
\item The Minkowski-space temperature $T(\tau,r)$ and its de Sitter analog $\hat{T}(\rho)$ are related by Weyl rescaling as 
\beq
T(\tau,r)=\frac{\hat{T}(\rho(\tau,r))}{\tau}\,. 
\eeq
For $\tau= 1.5$ fm/$c$ and $r=0$, Eq.~\eqref{definerho} gives us that $\rho\approx -1.2$. When using for the initial condition of the temperature in de Sitter space the natural scale $\hat{T}(\rho_0)=1$, we obtain from the numerical solution of Eq.~\eqref{eq:efftemp} that $\hat{T}(\rho=-1.2)=0.22$. Thus the corresponding central temperature $T(\tau= 1.5$ fm/$c$,~$r=0)=0.029$ GeV.
\end{itemize}
In order to visualize the distribution function in Minkowski space, we write 
\beq
\label{eq:func-Mink}
f(\tau, r, z=0,p_T,p_z)= f\left(\rho (\tau,r),\ph_\Omega^2 (\tau,r,p_T,p_z),\ph_\varsigma(\tau,p_z)\right)\,.
\eeq
and invert the functional dependence on the right hand side. At $\tau=1.5$ fm/$c$, and $z=p_\phi=0$ we find 
\begin{subequations}
\label{eq:Minkval}
\begin{align}
\label{eq:Minkvalr}
r&= 5\sqrt{-0.6\sinh\rho-0.91}\hspace{0.2cm}\text{fm}\,,\\
p_z&=0.13\,\, \ph_\varsigma\hspace{0.2cm} \text{GeV}\,,\\
\label{eq:Minkvalpt}
p_T&=
\begin{cases}
 0.13\,\gamma \left(\,\frac{\ph_\Omega}{\cosh\rho} + v\,\sqrt{\frac{\ph_\Omega^2}{\cosh^2\rho}+\ph^2_\varsigma}\right)\hspace{0.2cm} \text{GeV}& \text{if}\,\, p_T > \gamma v |p_z|\,,\\
 0.13\,\gamma \left(\,-\frac{\ph_\Omega}{\cosh\rho} + v\,\sqrt{\frac{\ph_\Omega^2}{\cosh^2\rho}+\ph^2_\varsigma}\right)\hspace{0.2cm} \text{GeV}& \text{if}\,\, p_T < \gamma v |p_z|
\end{cases}
\end{align}
\end{subequations}
where 
\beq
\begin{split}
v&=\frac{\sqrt{-0.6\sinh\rho-0.91}}{0.3-\sinh\rho}\,,\\
\gamma&=\frac{0.3-\sinh\rho}{\cosh\rho}
\end{split}
\eeq
Note that $r>0$ corresponds to $\rho<0$ at this value of $\tau$. 

\begin{figure}[t]
\begin{centering}
\includegraphics[scale=0.25]{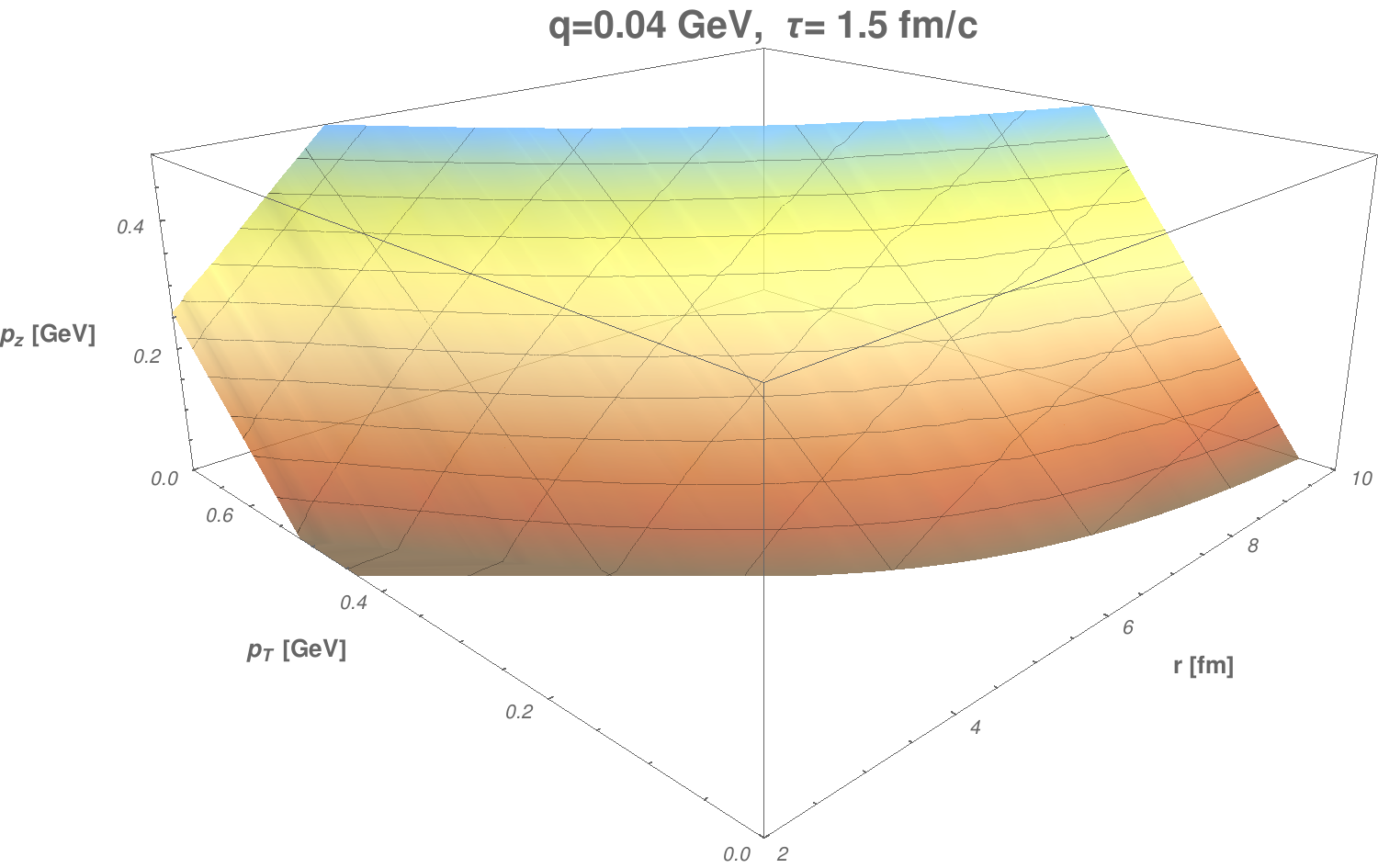}
\end{centering}
\caption{(Color online) 3D surface defined by $f(r,p_T,p_z)=0$ for $(4\pi)\eta/{\mathcal S}=3$ at $\tau=1.5$ fm/c and $q=0.04$ GeV. The gray lines drawn over the physical boundary condition $f=0$ correspond to constant values of $r-p_T-p_z$ over the surface.  See text for further details regarding the initial conditions.} 
\label{F6}
\end{figure}

In order to reconstruct the physical boundary in the $r-p_T-p_z$ phase space, we first determine numerically the set of coordinates $(\rho,\ph_\Omega,\ph_\varsigma)$ where $f=0$ for thermal initial conditions at $\rho_0=0$, using $(4\pi)\eta/S=3$ from Fig.~\ref{F5} (left panel). Once this information is determined in de Sitter space, it is simply mapped to Minkowski space by evaluating Eqs.~\eqref{eq:Minkval} for each selected data point $(\rho,\ph_\Omega,\ph_\varsigma)$. Note that the range of $(\rho,\ph_\Omega,\ph_\varsigma)$ covered in Fig.~\ref{F5} limits the range in $(r,p_T,p_z)$ that we can access in this way. To extend the $(r,p_T,p_z)$ range requires extending the numerical results shown in Fig.~\ref{F5} to a larger range in $(\rho,\ph_\Omega^2,\ph_\varsigma)$ which is numerically costly. For this reason we show in Fig.~\ref{F6} only a rather limited section of the surface in $(r,p_T,p_z)$ space that separates regions of positive and negative values for the distribution function $f$. 

In Fig.~\ref{F6}, the region where $f<0$ lies below the surface while the physical region $f>0$ lies above it. To determine the correct value for $p_T$ from \eqref{eq:Minkvalpt} we need to check the sign of $p_T-\gamma v |p_z|$. All points shown in Fig.~\ref{F6} correspond to $p_T>\gamma v p_z$, i.e. to the upper sign in Eq.~\eqref{eq:Minkvalpt}. We checked numerically that the opposite sign applies in regions of large $r$, beyond the range shown in Fig.~\ref{F6}. 

From Fig.~\ref{F6}, we observe that in the shown range $r\in [2\,\,\text{fm},10\,\,\text{fm}]$ the distribution function becomes negative for sufficiently large values of $p_T$ and sufficiently small values of $p_z$. As we move out to larger $r$ values (i.e. towards the tail of the density distribution) the unphysical region moves towards smaller $p_T$ values and thus covers a larger fraction of momentum space. For larger $r$, the solution with thermal boundary conditions at $\rho_0=0$ thus remains physical only for very small $p_T$ values combined with sufficiently large $|p_z|$. At fixed $r$, the critical surface separating $f>0$ from $f<0$ is almost linear in $p_T$ and $p_z$ and approximately given by $|p_z^{\mathrm{crit}}|=\alpha(r)(p_T-p_{T_0}(r))$ where $p_{T_0}(r)$ is the line where the critical surface intersects the $p_z=0$ plane. We find $\alpha(2\,\,\text{fm})\approx$ 1.47 and $\alpha(10\,\,\text{fm})\approx$ 2.08. These values depend on the initial choice of $\tau$ and all the other chosen parameters ($\hat{T}_0\equiv \hat{T}(\rho_0)$, $\eta/\mathcal{S}$ and $q$). For $p_z>|p_z^{\mathrm{crit}}|$ the solution is physical (i.e. $f>0$) while for $|p_z|<|p_z^{\mathrm{crit}}|$ it is unphysical ($f<0$).

\section{Conclusions}
\label{sec:concl}

In this work we have studied the dynamics in phase space of the recently found exact solution to the conformal RTA Boltzmann equation which undergoes Gubser expansion~\cite{Denicol:2014tha,Denicol:2014xca}. We determined the distribution function by first solving equation~\eqref{eq:efftemp} for the temperature $\hat{T}(\rho)$ and then  evaluating Eq.~\eqref{boltzmannsolution} point by point as a function of the variables $(\rho,\ph_\Omega,\ph_\varsigma)$ that define the phase space of the system in de Sitter space. We assumed thermal equilibrium boundary conditions at de Sitter time $\rho_0$. 

We observe that this exact solution becomes negative, and therefore physically meaningless, for sufficiently large negative values for $\rho-\rho_0$. This proves the conjecture in ~\cite{Denicol:2014xca} that the unphysical behaviour of the moments of the distribution function is likely related with the violation of positivity of the distribution function in certain phase space regions.

The non-physical behaviour of the distribution function depends strongly on the initial value  $\rho_0$. If the evolution of $f(\rho,\ph_\Omega,\ph_\varsigma)$ starts at $\rho_0=-\infty$ the system is always evolving in the forward direction in the de Sitter time $\rho$ and the distribution function remains always positive. When imposing thermal equilibrium initial conditions at some finite value $\rho_0$, the system can evolve not only forward but also backward in de Sitter time $\rho$. In this case the distribution function $f(\rho,\ph_\Omega,\ph_\varsigma)$ is always positive definite for $\rho-\rho_0 \geq 0$ but it loses its probabilistic and physical meaning in some regions of the phase space when $\rho-\rho_0 < 0$. Generically, the regions where the distribution function is negative occur at small $\ph_\varsigma$ and large $\ph_\Omega$,  qualitatively independent of the value of the shear viscosity over entropy density ratio $\eta/{\mathcal S}$ and the initial temperature $\hat{T}_0$. The generic shape of the surface separating the physical $(f>0)$ from the unphysical region is shown in de Sitter coordinates in Fig.~\ref{F5} and in Minkowski coordinates in Fig.~\ref{F6}. In Minkowski space, the exact solution for $f$ becomes unphysical at large $r$, large $p_T$ and small $p_z$; this corresponds to large negative values of $\rho-\rho_0$, large $\ph_\Omega$ and small $\ph_\varsigma$. The example studied in Sect.~\ref{sec:results} shows that when choosing values for the parameters $q$, $\hat{T}_0$ and $\eta/\mathcal{S}$ that are natural for heavy ion collisions, problems of non-positivity of $f$ arise already at moderately small values of $r$ and $p_T$ when $p_z=0$. This renders the analytical solution found in~\cite{Denicol:2014tha,Denicol:2014xca}, which assumes local thermal equilibrium on a surface $\rho_0$, unsuitable for heavy-ion phenomenology.
  
This does not mean, however, that this exact solution cannot be used to test different hydrodynamic approximation methods, as done in Refs.~\cite{Denicol:2014tha,Denicol:2014xca}. There is no problem with such tests as long as the comparison is performed (either in de Sitter or Minkowski space) in the physically allowed region where $f>0$. An alternative approach which guarantees always the positivity of the exact solution is to fix the initial condition at sufficiently large negative $\rho_0$ values such that the region $\rho >\rho_0$ includes all of the interesting range in $\tau$ and $r$ covered by the evolution of a heavy-ion collision. In this case, with a suitable choice for the initial form $f_0(\rho_0,\ph_\Omega,\ph_\varsigma)$ that is presumably not of equilibrium form, the exact solution of Eq.~\eqref{boltzmannsolution} might become relevant for heavy-ion phenomenology.

\acknowledgments
M. Martinez thanks D. Bazow, J. Jia and S. Koenig for their help with some of the numerical work. We thank J.~Noronha and G.~Denicol for useful discussions. This work was supported by the U.S. Department of Energy, Office of Science, Office of Nuclear Physics under Awards No. \rm{DE-SC0004286} and (within the framework of the JET Collaboration) \rm{DE-SC0004104}. 

\appendix

\bibliography{distfunc}

\end{document}